\documentclass[preprint,aps,nofootinbib]{revtex4}
\usepackage{graphicx}
\input{epsf.sty}
\usepackage{epsfig}

\newcommand{ \centeron }[2]{{\setbox0=\hbox{#1}\setbox1=\hbox{#2}\ifdim
                             \wd1>\wd0\kern.5\wd1\kern-.5\wd0\fi \copy0
                             \kern-.5\wd0\kern-.5\wd1\copy1\ifdim\wd0>\wd1
                             \kern.5\wd0\kern-.5\wd1\fi}}
\newcommand{ \ltap }{\>\centeron{\raise.35ex\hbox{$<$}}
                     {\lower.65ex\hbox{$\sim$}}\>}
\newcommand{ \gtap }{\>\centeron{\raise.35ex\hbox{$>$}}
                     {\lower.65ex\hbox{$\sim$}}\>}

\newcommand{ \slashchar }[1]{\setbox0=\hbox{$#1$}   % set a box for #1
   \dimen0=\wd0                                     % and get its size
   \setbox1=\hbox{/} \dimen1=\wd1                   % get size of /
   \ifdim\dimen0>\dimen1                            % #1 is bigger
      \rlap{\hbox to \dimen0{\hfil/\hfil}}          % so center / in box
      #1                                            % and print #1
   \else                                            % / is bigger
      \rlap{\hbox to \dimen1{\hfil$#1$\hfil}}       % so center #1
      /                                             % and print /
   \fi}                                             %

\def\singleandabitspaced{\baselineskip=\normalbaselineskip\multiply
    \baselineskip by 150\divide\baselineskip by 100}

\def\singlespaced{\baselineskip=\normalbaselineskip}

\begin{document}

\singlespaced

%\begin{titlepage}

\hfill$\vcenter{\hbox{\bf MADPH-05-1414}
                 \hbox{\bf hep-ph/0502102}}$
\vskip 0.4cm

\title{Remarks on Limits on String Scale from Proton Decay and Low-Energy amplitudes \\
in Braneworld Scenario } 
\author{ Piyabut Burikham \footnote{piyabut@physics.wisc.edu} }
\vspace*{0.5cm}
\affiliation{Department of Physics, University of Wisconsin, \\
        1150 University Avenue, Madison, WI 53706} 

\date{\today}

\vspace*{2.0cm}

\begin{abstract}
We discuss IR limit of four-fermion scattering amplitudes in braneworld models including intersecting-branes and SUSY $SU(5)$ GUT version of it.  With certain compactification where instanton effect is negligible, grand unification condition in D6-D6 intersecting-branes scenario subject to experimental constraint on proton decay provides possibility for upper limit on the string scale, $M_S$, through relationship between the string coupling, $g_s$, and the string scale.  We discuss how IR divergence is related to number of twisted fields we have to introduce into intersection region and how it can change IR behaviour of tree-level amplitudes in various intersecting-branes models.  Using number of twisted fields, we identify some intersecting-branes models whose tree-level amplitudes are purely stringy in nature and automatically proportional to $g_s/M^2_{S}$ at low energy.  They are consequently suppressed by the string scale.  For comparison, we also derive limit on the lower bound of the string scale from experimental constraint on proton decay induced from purely stringy contribution in the coincident-branes model, the limit is about $10^5$ TeV.      
\end{abstract}

\maketitle

%\end{titlepage}

\newpage

\setcounter{page}{2}
\renewcommand{\thefootnote}{\arabic{footnote}}
\setcounter{footnote}{0}
\singleandabitspaced

\section{Introduction}
\label{introduction-sec}

Proton decay has been an important issue which provides stringent test to various GUT models.  Conventional $SU(5)$ GUT, even being the simplest model, was ruled out by severe experimental limit on proton lifetime as well as its original SUSY version\cite{lan, his}(SGUT).  This is due to the dimension 5 proton decay in the $SU(5)$ SGUT model.  However, in models with extra dimensions, there are new ways to prevent proton decay e.g. by assuming nontrivial boundary condition on extra-dimensional components of fields\cite{hn, dei}.  Proton decay through dangerous dimension 5 operator could also be suppressed by the use of appropriate discrete symmetries\cite{dis}.  The leading contribution of proton decay is then of dimension 6 contact form being suppressed by the square of the mass scale.  With these developments, SGUT $SU(5)$ can be modified to survive experimental limit on proton lifetime.  

For SGUT $SU(5)$ in intersecting-branes models, symmetry is more naturally broken by discrete Wilson lines\cite{cshe}.  This is different from symmetry breaking mechanism in conventional 4-dimensional GUT.  We achieve gauge couplings unification by extra dimensional unification and it does not correspond, in general, to 4 dimensional GUT.  Threshold corrections in RGE in extra dimensional models contains extra contribution from massive Kaluza-Klein states.  This brings in dependence on geometrical factors, $L(Q)$, as well as volume of compactified manifold, $V_Q$.  They play the role of $M_{GUT}$ in the running of gauge couplings\cite{tor}.  In this sense, $M_{GUT}$ does not have any meaning in extra-dimensional unification but a parameter to keep track of unification expressed in 4 dimensional GUT language.   

In this paper, we will first discuss results from ref.~\cite{wit} on tree-level amplitudes in SUSY $SU(5)$ intersecting-branes model and the possibility of getting limit on the upper bound of the string scale in this D6-D6 model and proceed to discuss generic properties of quantum part of amplitudes in braneworld scenario in relation to number of twisted fields we introduce into the models.  Then we consider IR-correction to quantum part of amplitudes from classical-solutions contribution of the path integral, i.e. instanton contribution.  Quantum and classical contributions are discussed separately in order to emphasize unique characteristics of each one of them.  Phenomenology of braneworld scenario involves combination of effects from both local quantum behaviour and global classical contributions determined by compactification.  In this way we can discuss some possibilities that give purely stringy low energy amplitudes which do not have field theory correspondence.  One example of such processes could be proton decay as discussed in ref.~\cite{wit}.  Finally for comparison to "top-down" approach, we calculate proton decay in "bottom-up" coincident branes model using certain choices of Chan-Paton factors to kinematically suppress the amplitude\cite{jos}.  Estimated limit on lower bound of the string scale in this case is remarkably high.   
 
\section{IR-amplitudes in Intersecting-branes Models}

Generically, branes with any dimensionalities can intersect or be coincident.  There are a number of semi-realistic models of intersecting-branes with equal dimensionality\cite{shi} and therefore we will focus more on this case.  For completeness, we will also comment on IR behaviour of intersecting-branes with different dimensionalities such as D3-D7 configuration.  Finally we show that in certain situations, IR limit of string amplitudes in intersecting-branes scenario can be purely stringy with no standard model correspondence and they are automatically suppressed by the string scale.  

\subsection{Intersecting-branes with Equal Dimensionality and Proton Decay}

As in ref.\cite{wit}, we will consider dimension-6 channel of proton decay assuming dimension-5 channel is suppressed by some means such as discrete symmetries\cite{dis}.  In intersecting-branes model with particular $SU(5)$-group structure, leading contribution to proton decay is purely stringy \cite{wit}.  This is a dimension-6 operator proportional to string coupling $g_s$ and $\alpha^{\prime}=1/M^2_{S}$.  The formula for quantum amplitude of processes such as $p \to \pi^0e^+_L$ from ref.~\cite{wit} is
\begin{eqnarray}
A(1,2,3,4) & = & i\pi\frac{g_{s}}{M^2_S}I(\theta_{1}, \theta_{2}, \theta_{3})\bar{u_{1}}\gamma^{\mu}u_{2}\bar{u_{3}}\gamma_{\mu}u_{4}T_{1234}
\label{eq:1}
\end{eqnarray}
where
\begin{eqnarray}
I(\theta_{1}, \theta_{2}, \theta_{3}) & = & \int^1_{0}{\frac{dx}{x^{1+\alpha^{\prime}s}(1-x)^{1+\alpha^{\prime}t}}}\prod^3_{i=1}\frac{\sqrt{\sin{\pi\theta_i}}}{[F(\theta_{i}, 1-\theta_{i}; 1; x)F(\theta_{i}, 1-\theta_{i}; 1; 1-x)]^{1/2}},
\end{eqnarray}
$\theta_i$ are $SU(3)$ parameters relating 3 complex coordinates representing transverse directions to $1+3$ dimensional intersection region and $T_{1234}$ is corresponding Chan-Paton factor in $SU(5)$.  $F(x)\equiv F(\theta, 1-\theta; 1, x)$ is hypergeometric function.  Dependence on $F(x)$ comes from correlation function of four bosonic twisted fields 
\begin{eqnarray}  
<\sigma_{+}(0)\sigma_{-}(x)\sigma_{+}(1)\sigma_{-}(\infty)>& \sim & \sqrt{\sin{\pi \theta}}\frac{[x(1-x)]^{-2\Delta_{\sigma}}}{[F(x)F(1-x)]^{1/2}}
\label{eq:6}
\end{eqnarray}
with $\Delta_{\sigma}=\theta(1-\theta)/2$.
As $x\to 0$,
\begin{eqnarray}
F(x) \to 1, \   F(1-x)\to \frac{1}{\pi}\sin{\pi \theta}\ln(\frac{\delta}{x}) 
\label{eq:6.1}
\end{eqnarray}
where $\delta$ is some function of $\theta$ given in ref.\cite{abo}.  This asymptotic behaviour determines convergency of $x$-integration in the $s$-channel limit.    

In this setup, there is relationship between string parameters $(g_s, M_S, L(Q))$ and field theory GUT parameters $(\alpha_{GUT}, M_{GUT})$ as in Eq. (50) of \cite{wit},
\begin{eqnarray}   
g_s & = & \frac{\alpha_{GUT}L(Q)M^3_S}{(2\pi)^3M^3_{GUT}}
\label{eq:2}
\end{eqnarray}
where $L(Q)=4q\sin^2(5\pi w/q)$, Ray-Singer torsion, contains information on geometry of the compactified 3-manifold $Q=S^3/Z_{q}$\cite{wit, tor}.  This relationship relates $g_s$ to $M_S$ through numerical values of 4 dimensional $\alpha_{GUT}, M_{GUT}$.  Substitute this into Eq.~(\ref{eq:1}), we have
\begin{eqnarray}
A_{string} & = & i\frac{\alpha_{GUT}M_S}{2(2\pi)^2M^3_{GUT}}(L(Q)IT_{1234})\bar{u_{1}}\gamma^{\mu}u_{2}\bar{u_{3}}\gamma^{\mu}u_{4}
\end{eqnarray}
$I(s,t\to 0)$ is in $[7, 11.5]$ range, $L(Q)$ ranges from less than 1 to about order of 10.  With minimal choice that produces standard model gauges, $L(Q)=8$\cite{wit}.  Using numerical values of 4-dimensional $SU(5)$ SGUT(i.e. unification condition), $\alpha_{GUT}\simeq 0.04, M_{GUT}\simeq 2\times 10^{16}$ GeV leading to proton lifetime $\tau_{GUT}\simeq 1.6\times 10^{36}$ years\cite{his}, and experimental limit on proton lifetime, $\tau>4.4\times10^{33}$ years\cite{gan}, we have inequality
\begin{eqnarray}
\frac{\tau_{string}}{\tau_{GUT}} & = & \left| \frac{A_{GUT}}{A_{string}}\right|^2 > \left(\frac{4.4}{1.6}\right)\times 10^{-3}
\end{eqnarray}
leading to 
\begin{eqnarray}
M_S & < & 118M_{GUT}\simeq 2.4\times 10^{18}\mbox{ GeV} 
\label{eq:3}
\end{eqnarray}
where we have approximated $I\simeq 10$.  

There is also constraint from perturbative condition, $g_s<1$, using again Eq.~(\ref{eq:2}) with same set of numerical values, we have $M_S<9.2M_{GUT}\simeq 1.8\times 10^{17}$ GeV.  Grand unification and perturbative conditions together put limit on upper bound of string scale above which perturbative viewpoint breaks down.  Any SGUT(with D6-D6 configuration) string theories with larger $M_S$ would have to interact strongly and we need to consider proton decay in dual pictures.  The value of the upper bound of string scale, (\ref{eq:3}), is outside the perturbative constraint and therefore it is unfortunately inconclusive.  However, it is interesting that this upper limit on string scale does exist only in this D6-D6 model, if we have sufficiently more severe bound on proton decay in the future experiments, it would lead inevitably to limit on the upper bound of the string scale.   

An important aspect of this low-energy amplitude is the fact that it does not contain any $1/s$(Mandelstam's variable) pole like in conventional field theory amplitudes.  Rather it is proportional to $g_s/M^2_S$, we interpret this as a purely stringy effect which appears as contact interaction in field theory.  The advantage is it can suppress proton decay amplitude to be of the order of the string scale and therefore smallness is explained without the need of massive bosons exchange of the order of SGUT scale.  Remarkably, experimental limit on proton lifetime results in limit on UPPER bound of string scale in contrast to conventional SGUT cases where limit on lower bound of $X,Y$ bosons is derived.  Grand unification requirement in SGUT $SU(5)$ D6-D6 model relates string coupling to string scale and as a consequence, put limit on upper bound of the string scale.  

This result can be understood to be originated from difference between "top-down" and "bottom-up" approaches to string theory.  In top-down approach, we start with string parameters $(g_s, M_S)$ and geometrical details of compactification and we try to derive low energy parameters such as $g_{YM}, g1, g2, g3$, Yukawa coupling, mixing angles and so on.  With unification assumption, $g_s$ is tied to $M_S$ and geometrical factors and not a free parameter in the model.  Experimental constraint from proton decay then results in upper bound on $M_S$.  On the contrary, focussing mainly on kinematic extension of field-theory amplitudes to contain string resonances effect, bottom-up\cite{ks,pes,cor,jos,bhhm} approach simply fixes $g_s=g^2_{YM}$.  Without assuming unification, there is no particular relationship between $g_s$ and $M_S$.  This, in a traditional way, finally provides lower bound on the string scale when subject to experimental constraints\cite{pes, bhhm}.      

On the other hand, there seems to be disadvantage considering the need to have $1/s$ IR-divergence in order to reproduce field theory results at low energy\cite{bhhm}.  We need the correct IR limit of string amplitudes which contain the 0th mode pole as gauge boson exchange.  Intersecting-branes amplitudes actually provide $1/s$ pole in IR limit when we consider only one complex coordinate and one twisted field contribution together with classical contribution from two branes wrapping the same torus $T^2$\cite{abo, cp}.  Difference from the present case is due to differing number of twisted fields in the quantum part of the amplitude and the classical contribution of string winding modes which we will see later.  There are 3 sets of twisted-field(from 3 complex coordinates) correlation function, Eq.~(\ref{eq:6}), in the D6-D6 intersecting-branes model we are considering and they provide kinematic IR-regularisation to the amplitude\cite{wit}.  We can find critical number of twisted-field correlation functions above which IR divergence will be regularised by considering low energy expression for kinematic part of quantum amplitude containing $\ell$ twisted fields
\begin{eqnarray}            
\int^a_{0}\frac{dx}{x}(-\ln{x})^{-\ell /2} & = & \frac{2-\ell}{2}(-\ln{x})^{1-\ell/2}|^a_{0} \mbox{ for } \ell \neq 2 \\
                                        & = & - \infty \mbox{ for } \ell = 2
\label{eq:4}
\end{eqnarray}
where $a ~\epsilon~(0,1)$ is some small number, $\ell$ is number of correlation functions of twisted fields.
This is the same as Eq.~(22) in \cite{wit} when generalised to $\ell$ twisted fields.  The integration converges when $\ell\geq 3$ and therefore critical number of twisted fields is 3.  At least 3 twisted fields are required to regulate IR behaviour and this implies that we need to twist boundary condition of string in 3 complex coordinates of the model.  This is the case with D6-D6 setup.  

Using analytic continuation from negative $s$ to $s\geq 0$ like in usual Veneziano amplitude, we get some information on how the poles look like at $s=0$ and consequently at $s=nM^2_S$. 
\begin{eqnarray}
\int^a_{0}dx\frac{(-\ln{x})^{-\ell/2}}{x^{\alpha^{\prime}s+1}}& = & \int^{\infty}_{-\ln{a}}du~u^{-\ell/2}e^{\alpha^{\prime}su}
\label{eq:4.54}\\
                             & = & \frac{\Gamma (1-\frac{\ell}{2}, -s\ln{a})}{s^{1-\ell /2}}(\alpha^{\prime})^{\ell/2-1} 
\label{eq:4.55}
\end{eqnarray}
where incomplete Gamma function $\Gamma(x,y)\equiv \int^{\infty}_{y}e^{-u}u^{x-1}dx$.  Notably for $\ell = 2$, it gives $\sim (\ln{s})$ pole as $s\to 0$.  At $\ell =0$, we have normal gauge boson exchange $1/s$ pole.  For $\ell >0$, twisted fields modify pole by power of $\ell/2$.  At $\ell\geq 3$, amplitude is regulated.  Behaviour of all other poles at $s=nM^2_S$ for each value of $\ell$ are given by analytic continuation from pole at $s=0$ we have here.

In D$p$-D$p$ intersecting-branes($3<p<6$) with 1+3 dimensional intersection region, we need to change boundary condition of interbrane-attached string in $p-3$ complex dimensions.  Therefore we need to introduce $p-3$ twisted fields into each vertex operator(NS sector).  
Since the number of twisted fields is always less than 3, the amplitudes have IR divergences(not necessarily corresponding to gauge boson exchange) given by Eq.~(\ref{eq:4.55}).  In D5-D5, since $\ell=5-3=2$, quantum amplitude gives $(\ln{s})$ divergence.  In D4-D4, $\ell=1$ and we thus have fractional pole $1/s^{1/2}$.  In these models, we do not have purely stringy amplitudes, $g_s/M^2_S$, as leading order as in D6-D6 case.  However, from Eq.~(\ref{eq:4.55}), there are string resonance terms analytically continued from $s=0$ region.  At low energy, these terms $g_s {\alpha^{\prime}}^{\ell/2}/(s-nM^2_S)^{1-\ell/2}\simeq g_s/nM^2_S$.  Therefore there could be $g_s/M^2_S$ contact term in the amplitude regardless of the number of twisted fields $3>\ell >0$.  

Another curious aspect of amplitudes in SGUT intersecting-branes models is the factor $\alpha^{-1/3}_{GUT}$ enhancement comparing to 4 dimensional GUT amplitudes\cite{wit}.  We will see that this is the effect from compactification and it depends on how we achieve 1+3 world from 10 dimensional space.  Consider the parameters relation, Eq.~(\ref{eq:2}) could be generalized to D$p$-D$p$ case,
\begin{eqnarray}
g_s & \sim & \alpha_{GUT}\left(\frac{M^{p-3}_{S}}{M^{p-3}_{GUT}}\right)
\label{eq:4.59}
\end{eqnarray}
ignoring geometry factor.  This leads to 
\begin{eqnarray}
\frac{g_s}{M^2_S} & \sim & g_s^{1+2/(3-p)}\left(\frac{\alpha^{2/(p-3)}_{GUT}}{M^2_{GUT}}\right)
\label{eq:4.6}
\end{eqnarray}
which has enhancement factor $\alpha^{(5-p)/(p-3)}_{GUT}$ comparing to 4 dimensional GUT amplitude $\sim \alpha_{GUT}/M^2_{GUT}$.  Interestingly, this factor disappears at $p=5$ along with dependence of amplitude on $g_s$.  Since low energy limit of purely stringy part of tree-level amplitudes always appear as $g_s/M^2_s$ contact interaction form, we can conclude that stringy effect always appears with this enhancement(or dehancement) factor $\alpha^{(5-p)/(p-3)}_{GUT}$.  We can interpret the factor as a result from certain choice of compactification which gives our 1+3 dimensional matter universe.  Projected onto 4 dimensional field theory, fractional power of coupling $\alpha_{GUT}$ could as well be interpreted as "non-perturbative" characteristic of the amplitudes.  Observe also that coincident-branes limit $p=3 $ gives conventional "bottom-up" $g_s\sim \alpha_{GUT}$ identification and relationship between $g_s$ and $M_S$ remarkably disappears.  There is consistency between top-down and bottom-up approaches. 

\subsection{Intersecting-branes with Different Dimensionalities}

We can obtain $1+3$ intersection region from other combinations of intersecting-branes with differing dimensionalities.  An example of D3-D7 system has been calculated \cite{abl} and there is IR pole in the amplitude coming from instanton contributions cancelling effect of twisted fields as we will see later in section C.  Here we will focus only on quantum part of the amplitude and according to previous argument, we will show that IR behaviour is finite.  

Using again correlation function of four bosonic twisted fields, Eq.~(\ref{eq:6}), we reach at the same Eq.~(\ref{eq:4}) as a check for IR behaviour of the amplitude.  Since there are $\ell=4$ twisted fields in a vertex operator in order to change four boundary conditions of D7 to D3 which is larger than critical number of twisted fields (namely 3), therefore quantum part of four-fermion amplitude is finite and thus proportional to $g_s/M^2_S$ in low energy limit(from Eq.~(\ref{eq:4.55})).             

\subsection{Classical Contributions to String Amplitudes}

In path integral calculation of string scattering amplitude, the action is divided into quantum and classical contributions and they are factorized from one another.  Physically, quantum part depends only on local behaviour of quantum theory while classical part contains information of global geometry which constrains classical solutions of the system.  While classical contribution of path integral of field on sphere is constant and can be absorbed into string coupling(since there is no winding modes), classical contribution of field on nontrivial compactified manifold like torus contains various topological contributions from winding states.  We need these information to be manifest in order to extract correct low energy behaviour of string scattering amplitude.        

The simplest nontrivial case in which classical contribution has been calculated is $T^2$ torus with two branes wrapping specified by wrapping numbers $(n_1, m_1)$ and $(n_2, m_2)$\cite{abo,cp}.  Following ref.\cite{cp}, the classical contribution of the path integral is
\begin{eqnarray}  
\sum_{r_1,r_2}\exp-\frac{\sin(\pi\theta)}{2\pi\alpha^{\prime}}\left[\frac{F(1-x)}{F(x)}(r_{1}L_1)^2 + \frac{F(x)}{F(1-x)}(r_{2}L_2)^2\right]
\label{eq:10}
\end{eqnarray}
In the $x \to 0$ limit($s$-channel limit), the exponential contribution from $L_1$ lattice is zero except the zero mode, $r_{1}=0$ while the contribution from $L_2$ lattice becomes constant for each $r_2$.  With respect to one $r_2$ winding state, the contribution is just constant and low energy behaviour is thus governed totally by quantum part of the amplitude.  This would be the case if the winding states summation $\sum_{r_2}$ is somehow truncated at finite terms.  

However, we can use Poisson resummation to make some low energy behaviour manifest which can be seen explicitly in the Poisson resummation formula. 
\begin{eqnarray}
\sum_n \exp(-\pi an^2) & = &\sqrt{\frac{1}{a}}\sum_m \exp(-\frac{\pi}{a}m^2) 
\end{eqnarray}
The pole at $a=0$ arises on the left-handed side as an infinite sum of various instantons but not being manifest in each term.  The right-handed side manifests this pole as the volume factor in the upfront.  This pole becomes visible at low energy in this new "vacuum" choice after the Poisson resummation.  This resummation leads to
\begin{eqnarray}
\sum_{r_1,m_2}\sqrt{\frac{2\pi^2\alpha^{\prime}F(1-x)}{L^{2}_2\sin(\pi\theta)F(x)}}\exp-\frac{F(1-x)}{F(x)}\left[\frac{\sin(\pi\theta)}{2\pi\alpha^{\prime}}(r_{1}L_1)^2 + \frac{2\pi^3\alpha^{\prime}}{\sin(\pi\theta)}(\frac{m_{2}}{L_2})^2\right]
\label{eq:11}
\end{eqnarray}
for classical partition function.  The exponential of $F(1-x)/F(x)$ reduces to power of $x$ as $x\to 0$.
\begin{eqnarray}
\exp-\frac{F(1-x)}{F(x)}[...]& \sim & \left(\frac{\delta}{x}\right)^{-[...]\sin(\pi\theta)/\pi}
\label{eq:12}
\end{eqnarray}
This power of $1/x$ shift the $1/s$ pole as we can see from  
\begin{eqnarray}
[...]\frac{\sin(\pi\theta)}{\pi}& = & r^2_{1}\left(\frac{M^2_{S}}{M^2_{1}}\right)+\alpha^{\prime}(m_{2}M_{2})^2
\end{eqnarray}
where $M^2_{1}=2\pi^2/L^2_{1}\sin^2(\pi\theta), M^2_{2}=2\pi^2/L^2_{2}$ are corresponding KK masses.  With respect to $L_2$, the resonances appear at $s=(m_{2}M_{2})^2$.  With respect to $L_1$, the resonances appear at $s=r^2_{1}M^4_{S}/M^2_{1}$.  These are the usual KK and winding corrections which are not unexpected.  We can see that in the instanton-decoupled limit $M_S \gg M_c(M_c=1/L_1$ or $1/L_2$), the $r_1\neq 0$ contribution is very suppressed since the poles are at very high energies while the contribution from $M_2$ resonances are at low energies and thus non-negligible.  We can see that even each $r_2$ winding state contribution is suppressed, the infinite sum of their contributions become significant at low energies.  This is made manifest by Poisson resummation.   

Next we turn to the factor $\sqrt{F(1-x)/F(x)}$ in front of the exponential in Eq.~(\ref{eq:11}), this is the leading order contribution to $x$-integration of the amplitude and consequently the part that modifies effect of twisted fields to low energy physics.  Using approximation in Eq.~(\ref{eq:6.1}), the factor gives $(-\ln(x))^{1/2}$ as $x\to 0$.  This will modify the power of $-\ln(x)$ in Eq.~(\ref{eq:4.54}) to $(-\ln(x))^{(1-\ell)/2}$.  In other words, when there is one $T^2$, we replace $\ell$ by $\ell-1$, when there is two tori, $T^2\times T^2$, we replace by $\ell-2$ and so on.  We see that this piece results in fractional power of $1/s$, exotic kinematic effect which does not exist in field theory or KK models.  In D6-D6 model, we can assume two branes wrapping compactified space $T^2\times T^2\times T^2$\cite{cp}.  In this case, effects of twisted fields are completely compensated by these factors from classical contribution and we thus recover $1/s$ pole at low energy.  Therefore, around the resonances, since $x\simeq 0$ is dominant in the $x$-integration, the Veneziano form of the amplitude is naturally recovered in this choice of compactification.  Note that this is not necessary and there are possibilities for exotic IR behaviour, i.e. $g_s/M^2_S$ contact form or fractional power of $1/s$(Eq.~(\ref{eq:4.55})), of total amplitudes in other choices of compactification. 

The rule is if we have two intersecting branes wrapping same $n$ $T^2$ tori, we replace $\ell$ by $\ell-n$ in Eq.~(\ref{eq:4.55}) to get leading order behaviour of $1/s$ pole.  Complete cancellation occurs when $\ell=n$ and we always retrieve $1/s$ gauge boson exchange contribution.  In cases where $\ell-n>2$, we have IR finite amplitude and it is suppressed automatically by the string scale $M_S$ and effectively decouple at low energy.  In model construction, instead of arbitrary intersection and compactification choices(modulo previously known conditions such as SUSY preservation or GUT which are a matter of preferences), we also have to consider this kinematic aspect of string amplitudes.  For low energy phenomenology purpose, since we do not observe exotic fractional powers of $1/s$, therefore they should be eliminated by appropriate choices of compactification corresponding to number of twisted fields we have when we setup branes intersection.  At higher energy, there is no reasons(so far) to prevent these terms, they are part of stringy effects unique in intersecting-branes models.
        
Note also that after Poisson resummation, since the $1/s$ pole is recovered together with the factor of $M_2/M_S$ for each $T^2$( from Eq.~(\ref{eq:11})), the argument on the limit of the upper bound on the string scale from proton decay is no longer valid in this choice of compactification.  

On the other hand, instead of interpreting low energy physics in terms of field theoretic resonances(i.e. $x\to 0, 1$ limits corresponding to $s,t$-channel exchanges), it is pointed out in ref.~\cite{abo,fcnc,fcn} that there exists purely stringy contribution(instanton contribution) when contribution around saddle point of classical action is dominant in the $x$-integration.  However caution has to be made that this is the case only when quantum part of the amplitude is regulated(no singularity along $x$-integration).  If there is IR divergence from quantum part, it means the contribution from pole at $x=0(1)$ is dominant and saddle-point approximation ceases to be valid.  In the case that the quantum part is regulated, we can conclude from the previous section that leading order must be of contact form, $g_s/M^2_S$, now multiplying with exponential suppression from area of the worldsheet instanton.  As expected, even in this saddle-point approximation, the instanton effect is multiplied by $g_s/M^2_S$ and thus suppressed by the string scale.    

\section{Limit on lower bound of string scale in bottom-up approach from proton decay}

In "bottom-up" coincident-branes model\cite{pes,cor,bhhm}, we do not have effect of twisted fields in the picture, all fermions and gauge fields are identified with open string living on the same stack of branes with unspecified number of branes.  Assuming some unification group which have leptons and quarks in the same multiplet( in order to induce proton decay), we can identify each particle with appropriate Chan-Paton matrix.  Tree-level amplitude for 4 fermions is generically\cite{pes,bhhm,has}
\begin{eqnarray}
A_{string} & = & ig_{s} \left[A(s,t)S(s,t)T_{1234}+A(t,u)S(t,u)T_{1324}+A(u,s)S(u,s)T_{1243}\right]
\end{eqnarray}
where $A(x,y)$ is kinematic part of $SU(n)$ amplitude\cite{man,pes}, 
\begin{eqnarray}
S(x,y) & = & \frac{\Gamma(1-\alpha^{\prime}x)\Gamma(1-\alpha^{\prime}y)}{\Gamma(1-\alpha^{\prime}x-\alpha^{\prime}y)},
\end{eqnarray}
the usual part of Veneziano amplitude with the 0th pole excluded( put into $A(x,y)$ part explicitly).  Chan-Paton factors $T_{ijkl}=tr(t^it^jt^kt^l+reverse)$( $t$'s are Chan-Paton matrices) contains information of gauge group, mixing and so on of external particles.  To be more specific, we consider $u_{L}d_{R}\to \bar{u}_{R}e^{+}_{L}$ process of proton decay like in intersecting-branes case.  Proton decay amplitude is extremely small( if not 0) and therefore we match string amplitude with 0 at low energy.  Following ref.~\cite{bhhm}
\begin{eqnarray}  
A_{string}(f_{L}f_{R}\to f_{R}f_{L}) & = & ig_{s}\left[\frac{u^2}{st}T_{1234}S(s,t)+\frac{u}{t}T_{1324}S(t,u)+\frac{u}{s}T_{1243}S(u,s) \right]
\label{eq:7}\\
                                     & \simeq & 0        
\end{eqnarray}
where $s,t,u$ are conventional Mandelstam variables.  At low energy, $S(x,y)\to 1$ and since $s+t+u=0$, this leads to constraints on Chan-Paton factors, $T_{1234}=T_{1324}=T_{1243}\equiv T$.  Plug back into Eq.~(\ref{eq:7}), retrieving the next non-vanishing term from $S(x,y)\simeq 1-\frac{\pi^2}{6}\frac{xy}{M^4_S}$,
\begin{eqnarray}
A_{string}& = & -ig_{s}T\frac{\pi^2}{2}(\frac{u^2}{M^4_S})
\end{eqnarray}
Like in intersecting-branes case, we compare with $A_{GUT}$ and use experimental limit on proton decay while setting $T=1$( if $T=0$, there is no tree-level stringy proton decay and no limit on the string scale could be derived),
\begin{eqnarray}
\frac{\tau_{string}}{\tau_{GUT}} & = & \left| \frac{A_{GUT}}{A_{string}}\right|^2 = (\frac{4}{\pi^4})\frac{1}{u^2}(\frac{M^4_S}{M^2_{GUT}})^2 > \left(\frac{4.4}{1.6}\right)\times 10^{-3}
\end{eqnarray}
where we have identified $g_s=4\pi \alpha_{GUT}$.  At $E_{CM}\simeq 1$ GeV, $u\simeq 0.5$ GeV$^2$, this gives
\begin{eqnarray}
M_S & > & 8.5 \times 10^{7} \mbox{ GeV}\sim 10^{5} \mbox{ TeV}
\label{eq:8}
\end{eqnarray}
a remarkably strong limit on string scale.  Observe that this kind of kinematic suppression makes use of worldsheet duality( i.e. $s,t$ duality of Veneziano amplitude) to eliminate the contact interaction term $g_s/M^2_S$(dimension-6 operator), leaving only dimension-8 operator, $u^2/M^4_S$, as leading-order stringy correction which results in stringent limit on $M_S$.  This limit, however, ignores the conventional spontaneous symmetry breaking mechanism which suppresses proton decay by making the $X$ and $Y$ GUT bosons very massive.  It actually reflects the limit of proton decay from "purely stringy" effect which could exist if $T\neq 0$ in some specific embedding of the fermions in some unspecified open-string representation at higher energies.      
 
\section{Conclusions}
\label{conclusions-sec}

First we have discussed the possibility of getting limit on the upper bound of the string scale in D6-D6 intersecting-branes $SU(5)$ SGUT setup as in ref.~\cite{wit} from the experimental constraint on the proton decay.  The quantum part of the four-fermion tree-level amplitude in this case is of the contact form with $g_s/M^2_S$ dependence due to the number of twisted-field correlations.  We commented on how different number of twisted-field correlations in different D$p$-D$p$ setup could lead to different IR behaviour of the quantum part of the amplitude.   

Then we discussed appearance of the enhancement( or dehancement) factor $\alpha^{(5-p)/(p-3)}_{GUT}$ in D$p$-D$p$ setup when we compare stringy contact term $g_s/M^2_S$ to the $\alpha_{GUT}/M^2_{GUT}$ factor in 4 dimensional GUT amplitude.  This non-integer power of $\alpha_{GUT}$ is natural from the viewpoint that we "project" the extra-dimensional unification onto conventional 4 dimensional GUT RGE.  

In non-trivial compactification such as $T^2$, there are classical winding states contribution to the amplitude.  We explicitly demonstrated how Poisson resummation of the instanton contributions makes the classical instanton contribution to $x\to 0$ region manifest.  In intersecting-branes scenario, there are contributions from both quantum and classical part to the $x\to 0$ region in the stringy amplitude, and we need both to obtain the usual gauge boson $1/s$ pole at low energies. 

Finally we estimated the lower bound on the string scale in "bottom-up" coincident-branes approach using constraint on proton decay.  The limit is derived solely from purely stringy( of another kind of purely stringy effect from the dimension 6 mentioned above) contribution when appropriate choice of Chan-Paton factors is chosen.  Comparing to other constraints on the string scale in the "bottom-up" approach~\cite{bhhm}, this lower bound is remarkably strong, about $10^5$ TeV.      

\section*{Acknowledgments}
\indent
We wish to thank Tao Han, Gary Shiu, and Fernando Marchesano for valuable discussions.
This work was supported in part by the U.S. Department of Energy 
under grant number DE-FG02-95ER40896, and in part by the Wisconsin
Alumni Research Foundation.

%\vspace*{0.8cm}
%\newpage


\begin{thebibliography}{99}
\singlespaced

\bibitem{lan}
P. Langacker, Phys. \ Rep. \ {\bf 72C}, 185(1981).

\bibitem{his}
J. Hisano, hep-ph/0004266.

\bibitem{hn}
L. J. Hall and Y. Nomura, 
Phys.\ Rev.\ {\bf D65}, 125012 (2002)
[hep-ph/0111068].

\bibitem{dei}
K. R. Dienes, E. Dudas, and T. Gherghetta, 
Nucl.\ Phys.\ {\bf B537}, 47 (1999)
[hep-ph/9806292].

\bibitem{dis}
E. Witten, hep-ph/0201018.

\bibitem{cshe}
P. Candelas, G. Horowitz, A. Strominger, and E. Witten, 
 Nucl. \ Phys. \ {\bf B258}, 46(1985).

\bibitem{tor}
T. Friedmann and E. Witten, 
Adv.\ Theor.\ Math.\ Phys.\ {\bf 7}, 577 (2003)
[hep-th/0211269].

\bibitem{wit}
I. Klebanov and E. Witten, 
Nucl.\ Phys.\ {\bf B664}, 3 (2003)
[hep-th/0304079].

\bibitem{jos}
J. Friess, T. Han, and D. Hooper, Phys.\ Lett.\ {\bf B547}, 31 (2002)
[hep-ph/0204112].

\bibitem{shi}
G. Shiu and S.H. Tye,
Phys.\ Rev.\ {\bf D58}, 106007 (1998)\ [hep-th/9805157]; L.E. Ibanez, R. Rabadan, and A.M. Uranga, Nucl.\ Phys.\ {\bf B542}, 112 (1999)\ [hep-th/9808139]; I. Antoniadis, C. Bachas, and E. Dudas,
Nucl.\ Phys.\ {\bf B560}, 93 (1999)\ [hep-th/9906039]; K. Benakli, 
Phys.\ Rev.\ {\bf D60}, 104002 (1999)\ [hep-ph/9809582]; E. Accomando, I. Antoniadis and K. Benakli, Nucl.\ Phys.\ {\bf B579}, 3 (2000)\ [hep-ph/9912287]; M. Axenides, E. Floratos, and C. Kokorelis, JHEP {\bf 0310}, 006 (2003)\ [hep-th/0307255].

\bibitem{gan}
K. S. Ganezer,
Int.\ J.\ Mod.\ Phys.\ {\bf A16}, 855 (2001).

\bibitem{ks}
S. Kachru and E. Silverstein, 
Phys.\ Rev.\ Lett.\ {\bf 80}, 4855 (1998)
[hep-th/9802183].

\bibitem{pes}
S. Cullen, M. Perelstein, and M.E. Peskin, 
Phys.\ Rev.\ {\bf D62}, 055012 (2000)
[hep-ph/0001166].

\bibitem{cor}
F. Cornet, J.I. Illana, and M. Masip,
Phys.\ Rev.\ Lett.\ {\bf 86}, 4235 (2001)
[hep-ph/0102065].

\bibitem{bhhm}
P. Burikham, T. Han, F. Hussain, and D. McKay, 
Phys.\ Rev.\ {\bf D69}, 095001 (2004)
[hep-ph/0309132];
Piyabut Burikham, Terrance Figy, Tao Han,
Phys.\ Rev.\ {\bf D71}, 016005 (2005)
[hep-ph/0411094].

\bibitem{abo}
S. A. Abel and A. W. Owen, 
Nucl.\ Phys.\ {\bf B663}, 197 (2003)
[hep-th/0303124].

\bibitem{cp}
M. Cveti\v{c} and I. Papadimitriou, 
Phys.\ Rev.\ {\bf D68}, 046001 (2003); Erratum-ibid. {\bf D70}, 029903 (2004)
[hep-th/0303083].

\bibitem{abl}
I. Antoniadis, K. Benakli, and A. Laugier, 
JHEP {\bf 0105}, 044 (2001)
[hep-th/0011281]. 

\bibitem{fcnc}
S. A. Abel, M. Masip, and J. Santiago, 
JHEP {\bf 0304}, 057 (2003)
[hep-ph/0303087].

\bibitem{fcn}
S. A. Abel, O. Lebedev, and J. Santiago, 
Nucl.\ Phys.\ {\bf B696}, 141 (2004)
[hep-ph/0312157].

\bibitem{has}
A. Hashimoto and I.R. Klebanov,
Phys.\ Lett.\ {\bf B381}, 437 (1996)
[hep-th/9604065];
Nucl.\ Phys.\ Proc.\ Suppl.\ {\bf 55B}, 118 (1997)
[hep-th/9611214]. 

\bibitem{man}
M.L. Mangano and S.J. Parke,
Phys.\ Rep.\ {\bf 200}, 301 (1991).


\end{thebibliography}
\end{document}